\documentstyle[graphicx,float]{article}

\begin{document}
\title{Atomic absorption and emission in non-product states}
\date{}
\author{Pedro Sancho \\ Centro de L\'aseres Pulsados CLPU \\ Parque Cient\'{\i}fico, 37085 Villamayor, Salamanca, Spain}
\maketitle
\begin{abstract}
The properties of non-product states, both (anti)symmetrized and
entangled, have been extensively studied in the literature. In this
paper, by extending previous partial results, we introduce a general
framework to describe atomic absorption and emission in these
states. Our approach shows the existence of some fundamental
differences between entanglement and (anti)symmetrization and
between absorption and emission. For instance, modifications of the
emission properties with respect to that of uncorrelated systems
occur for any (anti)symmetrized state but only for some very
specific entangled states.
\end{abstract}

\section{Introduction}

Non-product states, entangled and (anti)symmetrized ones, are one of
the mainstream topics in physics. Effects associated with the
non-factorizable character of the states have been extensively
studied, specially violations of Bell's inequalities, teleportation
schemes and applications in quantum cryptography in the case of
entangled states, and detection of (anti)bunching and
Hong-Ou-Mandel-type interference in the realm of identical
particles. Another effects such as the influence of the
non-separability on the light-matter interaction and in particular
the optical properties of atomic systems, have not deserved so much
attention but have led to an independent line of research.

In this line, a review of the influence of entanglement in
collective states has been presented in \cite{pr}, and the
possibility of entanglement initiated super Raman scattering in
\cite{aga}. The modifications on the absorption properties of atoms
in (anti)symmetrized states have been described in \cite{ap}. A
specially relevant work is \cite{jap} where, up to our knowledge,
the only experimental contribution to the field is considered. The
authors measured some emission properties of excited atoms in
non-separable states showing that it differs from that of single
atoms (in product states). The lifetime of entangled excited atoms
has been studied in \cite{jpb}. The reverse of these phenomena
occurs when the atoms are radiated with light in non-factorizable
states. Some examples are \cite{bal} where two-photon absorption in
tapered optical fibers is enhanced, \cite{ros} whose authors studied
non-linear spectroscopy with entangled photons, and \cite{dow} a
review of the use of two-mode entangled states to improve the
precision of optical metrology. From an even more general
perspective, it has been shown the dependence of the light-matter
interaction on the separable or non-separable character of the
involved states. In particular, it has been demonstrated the
modification of the Kapitza-Dirac interference patterns when the state is (anti)symmetrized \cite{kdi} or entangled (also considering the case of identical particles) \cite{kde}. Finally, also related to our work, \cite{rev} provides a review of entanglement in atomic systems.

The aim of this paper is to present, based on previous partial
results, a general framework to study light absorption and emission
by atomic systems in non-product states. Our physical models mimic
those used in \cite{ap} and \cite{jpb} and consist of pairs of
atoms interacting with light fields. In order to simplify the
presentation we shall restrict our considerations to only two atoms.
The absorption and emission are described, in a phenomenological
way, as transitions between ground and excited states. In this
phenomenological model we do not explicitly consider the
electromagnetic field. The atoms are in non-product states, either
entangled or (anti)symmetrized ones. For entangled states the atoms
travel in opposite directions and are well-separated at the time
time of interaction with the light. A particular instance of this
model is the arrangement in \cite{jap}, where the photodissociation
of molecules leads to pairs of atoms moving in (entangled) almost
opposite directions. In contrast, for (anti)symmetrized states, the
identical atoms must be close in order to the exchange effects to be
important. The arrangements in both cases are very different and
will be treated separately. With these models we can determine general
expressions for absorption and emission. Moreover, our approach
illuminates the differences between entanglement and
(anti)symmetrization and between absorption and emission.

\section{Entangled states}

We study in this section the problem for entangled states. We
analyze separately the cases of absorption and emission.

\subsection{Absorption}

We consider a source of entangled pairs of distinguishable atoms
($A$ and $B$) traveling in opposite directions ($L$ and $R$). The
initial state is
\begin{equation}
|\psi _0>= \frac{1}{\sqrt 2}(|A>_L|B>_R + |B>_L|A>_R)
\end{equation}
When they are well-separated we introduce an interaction with a
light beam placed at $L$. We use a broad band frequency beam that
includes absorption frequencies for the two atoms. The light
intensity is low enough to discard double absorptions. This
procedure is equivalent to radiate simultaneously with two beams
sharped around the excitation frequencies of the two atoms.

A possible realization of this source is the dissociation of
diatomic molecules composed of two different isotopes of the same
species. The isotopes approximately match the above resonant
condition for the incident light. This source is reminiscent of the
arrangement in \cite{jap}, although in that case the two dissociated
atoms are identical and are in excited states.

After the interaction with the light the atoms evolve as $|A>_L
\rightarrow \alpha  |A^*>_L + \beta  |A>_L$ and $|B>_L \rightarrow
\gamma |B^*>_L + \delta  |B>_L$ with $A^*$ and $B^*$ denoting the
excited states of the atoms. The coefficients obey the normalization
conditions $|\alpha |^2 +|\beta |^2=1$ and $|\gamma |^2 +|\delta
|^2=1$. In order to avoid double absorptions we must have $|\alpha
_i|^2 \ll 1$ and $|\gamma _i|^2 \ll 1$. The state after the interaction is
\begin{eqnarray}
|\psi _i>=\frac{1}{\sqrt 2}( \alpha _L |A^*>_L |B>_R + \beta _L
|A>_L |B>_R + \nonumber \\ \gamma _L |B^*>_L |A>_R + \delta _L |B>_L
|A>_R)
\end{eqnarray}
The probability amplitudes of absorption by the atoms $A$ and $B$
are respectively $\alpha _L/\sqrt 2$ and $\gamma _L/\sqrt 2$. The
total probability of absorption in the state $\psi _i$ is obtained
by adding both probability amplitudes. In effect, the two absorption
alternatives (to be absorbed by atom $A$ or by atom $B$) are
indistinguishable as long as both are compatible with the broad band
range of the beam. The total one-absorption probability is
\begin{equation}
P_{one}=\frac{1}{2}|\alpha _L + \gamma _L|^2 = \frac{1}{2} |\alpha _L|^2 + \frac{1}{2} |\gamma _L|^2 + Re (\alpha _L^* \gamma _L)
\end{equation}
It must be compared with that associated with a mixture of the
product states $|A>_L|B>_R$ and $|B>_L|A>_R$ with equal weights
($1/2$), which is $P_{one}^{mix}=(|\alpha _L|^2 + |\gamma _L|^2)/2$.
Both expressions differ by the interference term $Re (\alpha _L^*
\gamma _L)$. It represents the interference effect associated with
the presence of indistinguishable absorption alternatives.
Entanglement modifies the absorption properties of the pair of
atoms.

At this point we must specify the observables associated with the
measurement of the absorption probability. In the one-particle case
that operator is the projector on state $A^*$, $\hat{\theta
}_A=|A^*><A^*|$, which acts as $\hat{\theta }_A|A^*>=|A^*>$ and
$\hat{\theta }_A|A>=0$, giving the eigenvalues $1$ and $0$
corresponding to excited and non-excited states. In the two-particle
case the operators are $\hat{O}_A=\hat{\theta }_A \otimes \hat{1}$
and $\hat{O}_B=\hat{\theta }_B \otimes \hat{1}$, acting
$\hat{\theta}$ on the left side and $\hat{1}$ on the right one. We
have $\hat{O}_A|A^*>_L|B>_R=|A^*>_L|B>_R, \cdots$. The observable
for the measurement of excited states at $L$ (independently of being
of type $A$ or $B$) is $\hat{O}_A + \hat{O}_B$. We would obtain the
same results using the non-local observable $ \hat{O}=\hat{\theta
}_A \otimes |B>_R<B| + \hat{\theta }_B \otimes |A>_R<A|$ (which is
the projector over the subspaces of excited states at $L$ and
non-excited ones at $R$). However, for the type of state and
interaction used in the paper the non-local observable would be
redundant from the local one. For instance, as at $R$ we must have
an atom of type $B$ if at $L$ there is one of type $A$ and the
interaction takes place at $L$, $|B>_R<B|$ in $\hat{O}$ is
equivalent to $\hat{1}$ in $\hat{O}_A + \hat{O}_B$. The fact that
the observable must be local can be seen from a possible
experimental realization of the measurement. We place photon
detectors at $L$, corresponding each detection event to the
spontaneous emission by an excited atom. There is not need for a
measurement process at $R$. In a more abstract form, the full
two-particle state reduces when the measurement at $L$ is performed,
being not necessary a measurement device at $R$. This is a
fundamental property of entangled states. There is a non-local
aspect in the measurement process, but it is related to the state
not to the observable.

To end the subsection we remark that this scheme is only valid for
distinguishable atoms. When the atoms are identical the initial state
reads as $|A>_L|A>_R$ that is not an entangled state but a
product one. As we shall see later, the absorption properties of
pairs of identical atoms can be modified, but due to (anti)symmetrization.

\subsection{Emission}

Let us consider a pair of atoms in the excited entangled state
\begin{equation}
|\psi _{exc}>= \frac{1}{\sqrt 2}(|A^*>_L|B>_R + |B^*>_L|A>_R)
\end{equation}
This state can be prepared, for instance, using the dissociation of
diatomic molecules with one of the atoms in an excited state. As in
subsection 2.1 this source closely remembers the arrangement
described in \cite{jap}.

The excited atoms spontaneously emit photons. We could expect
entanglement-based effects on the emission probabilities similar to
those found for absorption. However, the conditions for the
existence of these effects are now much more stringent. In effect,
the emission probabilities will give rise to interference effects
and consequently will differ from those of mixtures of product
states only when the photon emission is compatible with the two
alternatives (emission by $|A^*>_L|B>_R $ or by $|B^*>_L|A>_R $). In
general, different atoms have different emission frequencies,
becoming distinguishable the two alternatives. The two alternatives
are only indistinguishable in very particular circumstances, for
instance, when the emission frequencies are very close and cannot be
distinguished because of the radiative broadening. The simplest
realization of the scheme would be based on the use of two different
isotopes of the same species.

We evaluate the emission probability when the two alternatives are
indistinguishable. The description of the previous subsection is not
adequate for our problem because, for large times, all the excited
atoms decay to the ground state. We must instead consider emission
probabilities at a given time, that is, a time-dependent formalism.
Consequently a description based on the evolution operator $\hat{U}$
is more adequate for the problem. As after the preparation there is
not interaction between the atoms the evolution operator factorizes,
$\hat{U} = \hat{U}_A \hat{U}_B$. In the final state after emission
the two atoms are in the ground state and consequently, $|\psi
_f>=|\psi _0>$. Note that after the emission, as we have assumed the
photon to be compatible with the two alternatives, the superposition
is not broken and the state remains a pure one instead of becoming a
mixture. The matrix element for the emission transition is ${\cal
M}_{em}(t)=<\psi _f|\hat{U}(t)|\psi _{exc}>$. The emission
probability at time $t$, $ P_{em}(t)= |{\cal M}_{em}(t)|^2$, can be
written as
\begin{eqnarray}
P_{em}(t) =  \frac{1}{4} |{\cal M}_A^s(t){\cal M}_{B,nt}^s(t)|^2 +
\frac{1}{4} |{\cal M}_B^s(t){\cal M}_{A,nt}^s(t)|^2+ \nonumber \\
\frac{1}{2}Re({\cal M}_A^{s*}(t){\cal M}_{B,nt}^{s*}(t){\cal
M}_B^s(t) {\cal
M}_{A,nt}^{s}(t)) = \frac{1}{4} P_A^s(t) |{\cal M}_{B,nt}^s(t)|^2 + \nonumber \\
\frac{1}{4} P_B^s(t) |{\cal M}_{A,nt}^s(t)|^2+ \frac{1}{2} Re({\cal
M}_A^{s*}(t) {\cal M}_{B,nt}^{s*}(t) {\cal M}_B^s(t) {\cal
M}_{A,nt}^{s}(t))
\end{eqnarray}
with $ _L<A|\hat{U}_A|A^*>_L={\cal M}_A^s$ and
$_L<B|\hat{U}_B|B^*>_L={\cal M}_B^s$ the single-atom emission matrix
elements for $A$ and $B$. Their squared modulus are the single-atom
emission probabilities.  On the other hand, ${\cal
M}_{A,nt}^s=_R<A|\hat{U}_A|A>_R$ and ${\cal M}_{B,nt}^s
=_R<B|\hat{U}_B|B>_R$ are the non-transition matrix elements, which
only depend on the center of mass (CM) evolution of the state. To
derive the above equation we have used the fact that the terms
$_L<A|_R<B|\hat{U}|B^*>_L |A>_R$ and $_L<B|_R<A|\hat{U}|A^*>_L
|B>_R$ vanish.

This result must be compared with that for a mixture of initial
product states $|A^*>_L|B>_R$ and $|B^*>_L|A>_R$ with equal weights,
given by $P_{em}^{mix}(t)= (P_A^s(t) |{\cal M}_{B,nt}^s(t)|^2
+P_B^s(t) |{\cal M}_{A,nt}^s(t)|^2)/2$. We have again interference
effects between the two alternatives that lead to a modification of
the emission properties due to entanglement. We remark again that
for this scheme the effects only occur when the emission
alternatives are indistinguishable.

We represent graphically the above results. In the standard
description of spontaneous emission the total probability of
emission between the preparation of the state ($t=0$) and time $t$
is $P_A^s(t)=1-e^{-t/\tau _A}$ with $\tau _A$ the lifetime of atom
$A$. The matrix element can be expressed as ${\cal M}_A^s=M_A^s
<\psi _A|\hat{U}_{spa}(t)|\psi _{A^*}>$, with $M_A^s$ the matrix
element of the internal variables which must have the form
$M_A^s=(1-e^{-t/\tau _A})^{1/2}$ up to a phase factor, the $\psi$'s
the wave functions representing the spatial part of the states and
$\hat{U}_{spa}$ the spatial part of $\hat{U}$. $ <\psi
_A|\hat{U}_{spa}(t)|\psi _A^*>$ is the scalar product of the evolved
excited state and the non-excited one. If we assume that the recoil
associated with the emission and the spreading of the wave function
during the free evolution are small we have $ <\psi
_A|\hat{U}_{spa}(t)|\psi _{A^*}> \approx 1$ and we can take ${\cal
M}_A^s \approx M_A^s = (1-e^{-t/\tau _A})^{1/2}$. Similarly, as we
assume the spreading small we have ${\cal M}_{A,nt}^s \approx 1$.
With all these expressions, taking $\tau _A=1$ as the time unit and
$\tau _B=0.1$ we obtain the curves in Fig. 1. We see that at short
times the probability of emission is larger for product states.
\begin{figure}[H]
\center
\includegraphics[width=8cm,height=7cm]{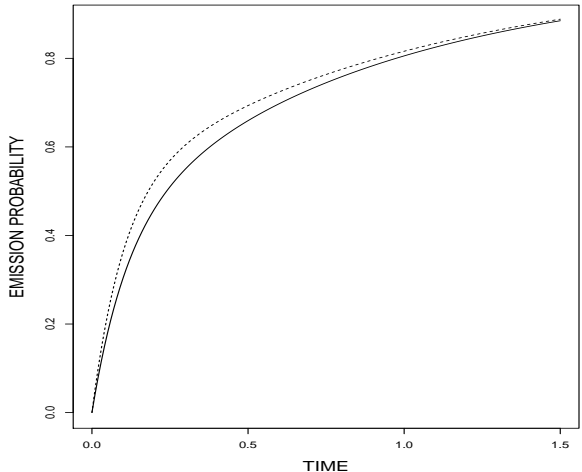}
\caption{Dependence of the emission probability on time (in units of
the lifetime of atom $A$). The continuous line represents entangled
states and the dashed one a mixture of product ones.}
\end{figure}

\section{(Anti)symmetrized states}

In this section we analyze the same problem for (anti)symmetrized
states of identical particles. The case of absorption has already
been considered in \cite{ap}. Here we shall extend that study to the
general case by discarding a simplification used there.
Later, we shall study emission.

A difference with the case of entangled states is that now, in
addition to the internal variables, we must explicitly introduce
into the formalism the CM-ones. In the calculations in the previous
section the CM-wave function only played a role similar to that of
the subscripts $L$ and $R$ labeling the alternatives. In contrast,
when the particles are identical the CM-wave functions determine the
overlapping of the two particles. The overlapping, defined as the
scalar product of the two wave functions, measures if the two wave
function domains lie in the same spatial region or in different
ones. Only in the first case the exchange effects are important.
The CM-wave functions must be included in all the steps of the
calculations as they even determine the normalization factors.

\subsection{Absorption}

We use the same notation introduced in \cite{ap}. The initial state of the
pair of identical atoms before the interaction with the light is
\begin{equation}
|\Psi _i>=N_i (|\phi _g>_1|\psi _g>_2 \pm |\psi _g>_1|\phi _g>_2)
\end{equation}
with the normalization factor $ N_i =(2( 1 \pm |<\phi | \psi >|^2))^{-1/2}$.
The state $|\psi _g> =|\psi >|g>$ is the product of the CM-wave
function and the internal state, denoting $g$ the ground state. The
upper sign in the double sign expressions refers to bosons and the
lower one to fermions.

When an atom absorbs a photon its state changes: $|\psi _g>
\rightarrow |\tilde{\psi }_e>$ and $|\phi _g> \rightarrow
|\tilde{\phi }_e>$, denoting $|e>$ the internal excited state.
Because of the recoil the CM-wave function changes to $\tilde{\psi
}$ or $\tilde{\phi }$. There are two alternatives for the absorption depending
on which atomic state the photon has been absorbed. These alternatives are represented by the states:
\begin{equation}
|\Psi _{abs} (\tilde{\psi})>=N_{abs}(\tilde{\psi}) (|\tilde{\psi }_e>_1|\phi _g>_2 \pm |\phi _g>_1|\tilde{\psi }_e>_2)
\label{eq:die}
\end{equation}
and
\begin{equation}
|\Psi _{abs} (\tilde{\phi})>=N_{abs}(\tilde{\phi}) (|\tilde{\phi }_e>_1|\psi _g>_2 \pm |\psi _g>_1|\tilde{\phi }_e>_2)
\label{eq:onc}
\end{equation}
with the normalization factor $ N_{abs}(\tilde{\psi})= (2( 1 \pm
|<\phi | \tilde{\psi } >|^2))^{-1/2}$. A similar expression holds
for $N_{abs} (\tilde{\phi})$. Note that, at variance with Eqs. (3)
and (4) of \cite{ap}, the normalization factors of Eqs.
(\ref{eq:die}) and (\ref{eq:onc}) are not $1/\sqrt 2$ but depend on
the overlapping. This difference is due to the fact that in
\cite{ap} it was assumed a small overlapping, equivalent to the
assumption of a large recoil, $|<\phi | \tilde{\psi } >|^2 \ll 1$
and $|<\psi | \tilde{\phi } >|^2 \ll 1$.

In \cite{ap}, in order to simplify the presentation, it was also
assumed that the final state after the absorption was a mixture.
This again corresponds to small overlapping. We do not longer rely
on this assumption and consequently we move to the general case. The
consequence of this extension is that now the two alternatives are
indistinguishable: as the overlapping is no longer negligible we
cannot distinguish if the photon was absorbed by the particle in
state $\psi $ or by that in $\phi $. We must add the probability
amplitudes. This is equivalent to consider a final pure state of the
system after the absorption, $|\Psi _f>=N_f (|\Psi
_{abs}(\tilde{\psi})>+|\Psi _{abs}(\tilde{\phi})>)$, with
$N_f=(2+2Re(<\Psi _{abs}(\tilde{\psi})|\Psi
_{abs}(\tilde{\phi})>))^{-1/2}$ and $<\Psi _{abs}(\tilde{\psi})|\Psi
_{abs} (\tilde{\phi})> =2N_{abs}(\tilde{\psi})
N_{abs}(\tilde{\phi})<\tilde{\psi}|\tilde{\phi}> <\phi |\psi> $,
where we have used $<e|g>=0$.

The one-absorption probability is $P_{one}^{ide}=|<\Psi _f
|\hat{U}|\Psi _i> |^2=N_f^2|{\cal M}_{\tilde{\psi}}+{\cal
M}_{\tilde{\phi}}|^2$ with the superscript $ide$ in the probability
denoting that now we are dealing with identical particles and ${\cal
M}_{\tilde{\psi}}=<\Psi _{abs} (\tilde{\psi})|\hat{U}|\Psi _i>$. We
fix the time $t$ at which the magnitudes are evaluated and we do not
longer include it explicitly into the equations. Assuming again that
there is no interaction between the atoms the evolution operator
factorizes, $\hat{U}=\hat{U}_1 \hat{U}_2 \equiv \hat{U}_s \hat{U}_s
$, with $\hat{U}_s$ denoting the single-particle operator. Using
this property it is simple to obtain ${\cal M}_{\tilde{\psi
}}=2N_{abs}(\tilde{\psi})N_i ({\cal M}_{\tilde{\psi} \phi}^s {\cal
M}_{\phi \psi}^s  \pm {\cal M}_{\tilde{\psi} \psi}^s {\cal M}_{\phi
\phi}^s )$ for the first alternative and ${\cal M}_{\tilde{\phi
}}=2N_{abs}(\tilde{\phi})N_i ({\cal M}_{\tilde{\phi} \phi}^s {\cal
M}_{\psi \psi}^s  \pm {\cal M}_{\tilde{\phi} \psi}^s {\cal M}_{\psi
\phi}^s )$ for the second one, denoting ${\cal M}_{\tilde{\psi}
\phi}^s =<\tilde{\psi}_e|\hat{U}_s|\phi _g>$, . .. the
single-particle matrix elements.

From the above equations it is possible to analyze the mathematical
structure of the absorption probabilities. There are three physical
factors that determine the form of the probabilities: the
overlapping of the CM-wave functions present in the normalization
coefficients, the indistinguishability of the absorption
alternatives ruling the addition of probability amplitudes, and the
(anti)symmetrized character of the wave function. The last two
factors give rise to interference effects. To study its form it is
convenient to express the two-particle matrix elements as ${\cal
M}_{\tilde{\psi}}= {\cal M}_{\tilde{\psi}}^d \pm {\cal
M}_{\tilde{\psi}}^e$,.. denoting the superscripts $d$ and $e$ the
direct and exchange terms. With this notation the absorption
probability reads
\begin{eqnarray}
P_{one}^{ide}/N_f^2= |{\cal M}_{\tilde{\phi }}^d|^2 + |{\cal M}_{\tilde{\phi }}^e|^2 \pm  2Re({\cal M}_{\tilde{\phi }}^{d *} {\cal M}_{\tilde{\phi }}^e ) + |{\cal M}_ {\tilde{\psi }}^d|^2 + |{\cal M}_{\tilde{\psi }}^e|^2 \pm \nonumber \\
2Re({\cal M}_{\tilde{\psi }}^{d *} {\cal M}_{\tilde{\psi }}^e ) +
2Re({\cal M}_{\tilde{\phi }}^{d *} {\cal M}_{\tilde{\psi }}^d \pm
{\cal M}_{\tilde{\phi }}^{d *} {\cal M}_{\tilde{\psi }}^e \pm {\cal
M}_{\tilde{\phi }}^{e *} {\cal M}_{\tilde{\psi }}^d + {\cal
M}_{\tilde{\phi }}^{e *} {\cal M}_{\tilde{\psi }}^e )
\end{eqnarray}
The four squared modulus terms represent the probabilities
associated with the four absorption alternatives available to the
system: $(\phi _g \rightarrow \tilde{\psi }_e ; \psi _g \rightarrow
\phi _g)$, $(\psi _g \rightarrow \tilde{\psi }_e ; \phi _g
\rightarrow \phi _g)$, $(\psi _g \rightarrow \tilde{\phi }_e ; \phi
_g \rightarrow \psi _g)$ and $(\phi _g \rightarrow \tilde{\phi }_e ;
\psi _g \rightarrow \psi _g)$, corresponding respectively to ${\cal
M}_{\tilde{\psi }}^d , {\cal M}_{\tilde{\psi }}^e , {\cal
M}_{\tilde{\phi }}^d $ and ${\cal M}_{\tilde{\phi }}^e $.

The other terms correspond to the six interference effects existent
between these four alternatives. We must distinguish between three
different types of interference effects: exchange,
indistiguishability and mixed ones. The two exchange terms (${\cal
M}_{\tilde{\phi }}^{d *} {\cal M}_{\tilde{\phi }}^e $ and ${\cal
M}_{\tilde{\psi }}^{d *} {\cal M}_{\tilde{\psi }}^e $) emerge
directly from the superposition of direct and exchange terms with
the same final states ($\tilde{\phi}_e , \psi _g$ and $\tilde{\psi
}_e , \phi _g$). They differ for bosons and fermions via the
characteristic double sign $\pm$. On the other hand, the
indistinguishability terms (${\cal M}_{\tilde{\phi }}^{d *} {\cal
M}_{\tilde{\psi }}^d $ and ${\cal M}_{\tilde{\phi }}^{e *} {\cal
M}_{\tilde{\psi }}^e $) are equal for bosons and fermions and their
matrix element products only contain direct or exchange terms. Their
physical origin is the existence of indistinguishable alternatives (final states
$\tilde{\phi}_e , \psi _g$ and $\tilde{\psi }_e , \phi _g$ after the
absorption). Finally, the mixed terms (${\cal M}_{\tilde{\phi }}^{d
*} {\cal M}_{\tilde{\psi }}^e $ and ${\cal M}_{\tilde{\phi }}^{e *}
{\cal M}_{\tilde{\psi }}^d $) share characteristics with the
previous ones. They contain direct and exchange matrix elements reflecting an
exchange-type contribution, but at the same time they refer to
different final states, a typical property of
indistinguishability-type terms. Then both types of physical effects
are present in the mixed terms. In particular, they differ for
bosons and fermions.

We represent graphically the absorption probabilities. As discussed
for Fig. 1 the matrix elements ${\cal M}_{\tilde{\psi} \phi}^s$
represent the product of the internal matrix element and the
overlapping of $\tilde{\psi}$ and the evolved $\phi (t)$. We assume
again that the recoil of the atom by the absorption is small and
that the spreading of $\phi$ during the free evolution is not too
large (taking for $t$ small values after the atom-light
interaction). Then we can approximate the spatial part of ${\cal
M}_{\tilde{\psi} \phi}^s$ by the time-independent overlapping
$<\tilde{\psi}| \phi>$. We take for the internal matrix element
$M^s=1$ and represent the absorption probability versus the initial
overlapping $<\psi |\phi >$. We use for the other overlapping
parameters the values $<\tilde{\psi} |\tilde{\phi }>=0.5$ and $0.9$,
$<\tilde{\psi }|\phi >=0.6=<\tilde{\phi }|\phi >$, $<\psi
|\tilde{\phi }>= <\psi|\phi ><\tilde{\psi}|\phi >$ and $<\tilde{\psi
}|\psi > =<\psi |\phi ><\tilde{\psi}|\phi >$ (choice equivalent to
fix $\phi$,$\tilde{\phi}$ and $\tilde{\psi}$ and to vary $\psi$).
\begin{figure}[H]
\center
\includegraphics[width=7cm,height=7cm]{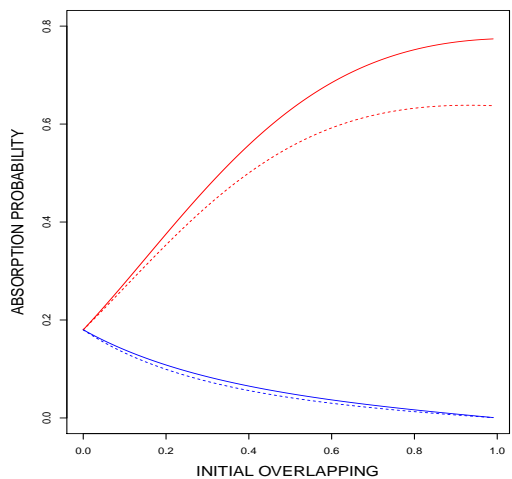}
\caption{Variation of the absorption probability with respect to the
initial overlapping. The red and blue curves correspond respectively
to bosons and fermions, and the continuous and discontinuous ones to the values $0.5$ and $0.9$ of $<\tilde{\psi} |\tilde{\phi }>$.}
\end{figure}
The results are presented in Fig. 2. For zero initial overlapping we
have the absorption probability of distinguishable particles. When
that overlapping increases we observe deviations from the
distinguishable behavior. The probabilities for fermions are always
smaller. In particular, when the initial overlapping tends to one
the probability vanish, reflecting the Pauli exclusion principle. In
contrast, for bosons the probabilities always increase with respect
to the values of non-identical particles. The variation between the
curves for different values of $<\tilde{\psi} |\tilde{\phi }>$ is
much larger in the case of bosons.

\subsection{Emission}

We consider now emission by excited states of identical particles.
For the sake of concreteness we only consider the case with the two
atoms initially excited. This is the most interesting case because
it is closely related to the experimental arrangement in \cite{jap},
where the photodissociation of a $H_2$ molecule leads to two excited
Hydrogen atoms. The initial state is
\begin{equation}
|\Psi _{exc}> = N_i (|\phi _e>_1|\psi _e>_2 \pm |\psi _e>_1|\phi _e>_2)
\end{equation}
We want to evaluate the probability of double emission after a fixed
time that, as signaled before, is not explicitly included in the
equations. Note that the direction of photon emission is random and
varies from repetition to repetition of the experiment. After the
emission the atom suffers a recoil and its CM-wave function also
changes in a random way. We make the evaluation for only two fixed
emission directions, which correspond to the final state
\begin{equation}
|\Psi _F(\overline{\phi },\overline{\psi })> = N_F (\overline{\phi },\overline{\psi }) (|\overline{\phi }_g>_1|\overline{\psi }_g>_2 \pm |\overline{\psi }_g>_1|\overline{\phi }_g>_2)
\end{equation}
with $ N_F (\overline{\phi },\overline{\psi })= (2( 1 \pm
|<\overline{\phi} | \overline{\psi } >|^2))^{-1/2}$. Note that as
the emission processes are indistinguishable (the particles are
identical) the final state is a pure one. This property agrees with
the fact that the final state of the atom-photon system after
emission is a pure entangled one \cite{wei,rza}. In our case, as the
light field variables are not explicitly included, the total
atom-photon pure state can be expressed in the form $\Psi
_F(\overline{\phi },\overline{\psi })$.

The probability amplitude for the transition is $<\Psi
_F(\overline{\phi },\overline{\psi })|\hat{U}|\Psi _{exc}>$,
and the probability can be written as
\begin{equation}
P_{two}^{ide}=\frac{P_{\overline{\phi } \phi }^s P_{\overline{\psi }
\psi }^s  + P_{\overline{\phi } \psi }^s P_{\overline{\psi } \phi
}^s \pm 2Re ( {\cal M}_{ \overline{\phi } \phi  }^{s*} {\cal M}_{
\overline{\psi } \psi  }^{s*} {\cal M}_{ \overline{\phi } \psi }^{s}
{\cal M}_{ \overline{\psi } \phi  }^{s}  )}{( 1 \pm |<\phi | \psi
>|^2) ( 1 \pm |<\overline{\phi} | \overline{\psi } >|^2)}
\end{equation}
where, as in the previous section, $P_{\overline{\phi}
\phi}^s=|{\cal M}_{\overline{\phi} \phi}^s |^2$ and ${\cal
M}_{\overline{\phi} \phi}^s= <\overline{\phi}_g|\hat{U}_s|\phi _e>$.

From this equation we see that the emission probability depends on
the overlapping between both the initial ($<\phi | \psi >$) and the
final ($<\overline{\phi} | \overline{\psi }>$) CM-wave functions and
on the exchange interference between the two available alternatives,
$(\phi _e \rightarrow \overline{\phi }_g ; \psi _e \rightarrow
\overline{\psi }_g  )$ and $(\psi _e \rightarrow \overline{\phi }_g
; \phi _e \rightarrow \overline{\psi }_g  )$.

\section{Discussion}

We have analyzed the absorption and emission processes in
non-product atomic states in two archetypical situations that
represent a rather general framework for the problem. The fundamental
characteristics of most arrangements can be described via these
situations with minor changes. Some properties of these processes in
non-factorizable states have been previously presented in the
literature but in a somewhat disperse way. We consider useful the
unified view introduced here.

The most important consequence of our analysis is to show that the
absorption and emission probabilities in non-factorizable atomic
states differ in many cases with respect to those of product states.
This property can be seen as a distinctive characteristic of
non-product states, just as violations of Bell's inequalities,
(anti)bunching,... These differences have a rich and diverse
structure. The most extreme case is that of absorption in
(anti)symmetrized states, where four types of modifications are
present: dependence on the overlapping and three types of
interferences, by exchange effects, by the existence of
indistinguishable alternatives and by the joint action of the two
previous causes.

In addition to provide a general framework for the problem, our
approach illustrates some aspects of the subject that deserve
attention. The first one is the relation between entanglement and
(anti)symmetrization. There is some controversy in the literature
about this point. Some times they are considered as rather similar
concepts because both represent multi-particle superpositions. As a
matter of fact, from a mathematical point of view, they are
described by the same type of non-factorizable equation (with the
important difference of the normalization coefficient). However,
there are some fundamental differences between both types of
superpositions. The first one, and most evident, is that exchange
effects only occur when the particles are very close whereas
entanglement-based effects can be present at large separations (as
in the arrangement here considered). This property is closely
related to the different role of the CM-wave function in both cases.
For identical particles it determines the strength of the exchange
effects and its scalar product enters into the normalization factor.
In contrast, for entangled systems it plays a secondary role
equivalent to that of the labels that denote if one particle goes to
the left or to the right. In this paper we have shown another
important difference: (anti)symmetrization modifies the emission
properties for any excited state of the multi-particle system (even
with the atoms in different excited states) whereas entanglement can
only do it when the emission alternatives are indistinguishable. For
identical particles it is an universal property and for entangled
ones almost an exception.

All the considerations in this paper about identical particles refer
to states without entanglement, that is, the non-separability in the
system is exclusively associated with the (anti)symmetrization. A
natural further step is to study identical particles in entangled
states (see \cite{rev} for a general view and \cite{kde,tic,iem,gig}
for particular considerations). This is an interesting and
controversial field. In particular, with reference to the
fundamental role played by the overlapping in systems of identical
particles, the authors of \cite{tic} have analyzed not only the
overlapping of the states (as we have done here), but also how the
overlapping evolves during the process of interaction with the
detectors, quantifying the mutual dependence between measurement
processes and effective particle distinguishability. On the other
hand, in \cite{iem} a more general framework for the problem, based
on the quantumness of correlations, has been introduced. Finally, in
\cite{gig} entanglement measures for fermionic systems with not
fixed particle number have been discussed.

Another point illustrated by our analysis refers to the differences
between absorption and emission. As discussed before the emission
properties are radically different for (anti)symmetrized and
entangled states, whereas such a sharp contrast is not present for
absorption. Another manifestation only concerning to entangled
states, is that in the presence of entanglement the absorption rates
are modified with respect to those of product states, in marked
contrast with emission where the process only occurs when the
emission alternatives are indistinguishable. The modification of the
absorption properties is a much more general phenomenon than in the
case of emission.

\end{document}